# Quantum steganography with large payload based on dense coding and entanglement swapping of GHZ states

Tian-Yu Ye*, Li-Zhen Jiang

E-mail：yetianyu@zjgsu.edu.cn

College of Information & Electronic Engineering, Zhejiang Gongshang University, Hangzhou 310018, P.R.China

**Abstract**

A quantum steganography protocol with large payload is proposed, based on dense coding and entanglement swapping of GHZ states. Its super quantum channel is formed by building up the hidden channel within the original quantum secure direct communication (QSDC) scheme. Based on the original QSDC, secrets messages are transmitted by integrating dense coding of GHZ states and entanglement swapping of GHZ states. Its capacity of super quantum channel achieves six bits per round covert communication, much higher than the previous quantum steganography protocols. Its imperceptibility is good, since the information and secrets messages can be regarded to be random or pseudo-random. Moreover, its security is proved to be reliable.

## 1 Introduction

Quantum secure communication can transmit secret messages with unconditional security by using the law of quantum mechanics, and can be classified into quantum key distribution (QKD)[1-4], quantum secure direct communication (QSDC) [5-14], quantum secret sharing (QSS) [15-18] and so on. The goal of QKD is to establish a shared secret key between two remote authorized users. In 1984, Bennett and Brassard[1] proposed the famous BB84 protocol, which is the first QKD protocol. Afterward, a lot of QKD schemes have been put forward[2-4]. QSDC allows secret messages to be transmitted directly without creating a key to encrypt them in advance. In 2002, Beige et al.[5] put forward the first QSDC protocol. On the other hand, Bostrom and Felbinger[6] suggested the famous Ping-Pong protocol in 2002. In 2003, Deng et al.[7] presented a two-step QSDC based on Bell states. In 2004, Cai and Li[8] introduced two additional unitary operations to improve the capacity of Ping-Pong protocol[6]. In 2005, Wang et al.[9] put forward a multi-step QSDC using multi-particle GHZ state. In 2005, Gao et al.[10] put forward a simultaneous QSDC between the central party and other M parties. In 2008, Sun et al.[11] proposed a new multiparty simultaneous QSDC based on GHZ states and dense coding. In 2008, Chen et al.[12] proposed a novel controlled QSDC protocol with quantum encryption using a partially entangled GHZ state. In 2008, Chen et al.[13] proposed a novel three-party controlled QSDC protocol based on W state. In 2012, Huang et al.[14] proposed two robust channel-encrypting QSDC protocols over different collective-noise channels. As the quantum counterpart of classical secret sharing, QSS allows that neither individual agent is able to obtain secret messages sent by the dealer unless they cooperate. In 1999, Hillery et al.[15] put forward the first QSS protocol with GHZ state. In 2003, Guo et al.[16] presented a QSS protocol only using product states. In 2010, Yang et al.[17] put forward a novel three-party QSS protocol of secure direct communication based on $\chi$-type entangled states. In 2012, Chen et al.[18] put forward a three-party QSS protocol via the entangled GHZ state using the single-particle quantum state to encode the information.

Classical steganography inserts secret messages into an innocent cover object to hide the existence of secret messages. In recent years, the concept of quantum steganography has been put forward, which applies the concept of classical steganography into a quantum scenario. Different from the traditional quantum secure communication, quantum steganography always builds up another hidden channel within normal quantum channel to transmit secret message. That is, quantum steganography uses the hidden channel to hide the existence of secret messages into normal quantum channel. In 2002, Gea-Banacloche[19] used quantum error-correcting code(QECC) to encode quantum data and hided secret messages as errors. In 2004, Worley III[20] suggested a fuzzy quantum watermarking protocol with the relative frequency of error from observing qubits. In 2007, based on the BB84 QKD scheme[1], Martin[21] presented a novel quantum steganography protocol. In 2010, based on Guo et al.'s QSS scheme[16], Liao et al.[22] proposed a novel multi-party quantum steganography protocol. In 2010, based on improved ping-pong scheme (IBF)[8], Qu et al.[23] suggested a novel quantum steganography scheme with large payload. However, the capacity of quantum channel in Ref.[19], Ref.[20], Ref.[21] and Ref.[22] is only one bit or one qubit per round covert communication, which is apparent to be too small to efficient covert communication. Although the capacity in Ref.[23] has been increased to four bits, it is still not large enough.

Based on the above analysis, for improving quantum channel's capacity, we propose a novel quantum steganography protocol with large payload based on dense coding and entanglement swapping of GHZ states. Its super quantum channel is formed by building up the hidden channel within the original quantum secure direct communication (QSDC) scheme. Based on the original QSDC, secrets messages are transmitted by integrating dense coding of GHZ states and entanglement swapping of

GHZ states. It can transmit six bits per round covert communication, six times that of Ref.[19], Ref.[20], Ref.[21]or Ref.[22]and one and half times that of Ref.[23]. Since the information and secrets messages can be regarded to be random or pseudo-random, its imperceptibility is good. Moreover, its security is proved to be reliable.

## 2  Coding scheme

The dense coding of GHZ states is briefly introduced firstly. The dense coding of GHZ states, proposed by Lee et al.[24], is generalization of the dense coding scheme of Bennett and Wiesner[25] in the case of GHZ states. GHZ-states are three-particle maximally entangled states, and form a complete orthogonal basis of 8-dimensional Hilbert space. The eight independent GHZ-states are expressed as follows:

$$|\Psi_1\rangle = \frac{1}{\sqrt{2}}(|000\rangle+|111\rangle), |\Psi_2\rangle = \frac{1}{\sqrt{2}}(|000\rangle-|111\rangle), |\Psi_3\rangle = \frac{1}{\sqrt{2}}(|100\rangle+|011\rangle), |\Psi_4\rangle = \frac{1}{\sqrt{2}}(|100\rangle-|011\rangle),$$

$$|\Psi_5\rangle = \frac{1}{\sqrt{2}}(|010\rangle+|101\rangle), |\Psi_6\rangle = \frac{1}{\sqrt{2}}(|010\rangle-|101\rangle), |\Psi_7\rangle = \frac{1}{\sqrt{2}}(|110\rangle+|001\rangle), |\Psi_8\rangle = \frac{1}{\sqrt{2}}(|110\rangle-|001\rangle). \quad (1)$$

By performing single-particle unitary operations on any two of the three particles, one GHZ-state can be transformed into another GHZ-state, where four single-particle unitary operations are

$$I = |0\rangle\langle 0| + |1\rangle\langle 1|, \sigma_z = |0\rangle\langle 0| - |1\rangle\langle 1|, \sigma_x = |0\rangle\langle 1| + |1\rangle\langle 0|, i\sigma_y = |0\rangle\langle 1| - |1\rangle\langle 0|. \quad (2)$$

Without loss of generality, $|\Psi_1\rangle$ is assumed as the initial state. Accordingly, $|\Psi_1\rangle$ can be transformed into $|\Psi_k\rangle$ ($k=1,2,\cdots,8$) by performing $U_k$ on its first and second particle, namely,

$$U_k|\Psi_1\rangle = |\Psi_k\rangle \ (k=1,2,\cdots,8), \quad (3)$$

where

$$U_1 = \sigma_z \otimes \sigma_z, U_2 = I \otimes \sigma_z, U_3 = i\sigma_y \otimes \sigma_z, U_4 = \sigma_x \otimes \sigma_z, U_5 = I \otimes \sigma_x, U_6 = \sigma_z \otimes \sigma_x, U_7 = \sigma_x \otimes \sigma_x, U_8 = i\sigma_y \otimes \sigma_x. \quad (4)$$

Let each $U_k$ correspond to three bits information, namely,

$$U_1 \to 000, U_2 \to 001, U_3 \to 010, U_4 \to 011, U_5 \to 100, U_6 \to 101, U_7 \to 110, U_8 \to 111. \quad (5)$$

Based on the above description, after the dense coding of GHZ states, one GHZ state can transmit three bits information.

The results of entanglement swapping between $|\Psi_1\rangle$ and arbitrary one of eight GHZ-states are shown as formulas (6)-(13), where the subscript $A_i$, $B_i$, $C_i$ ($i=1,2$) denote three particles in GHZ states, respectively.

$$|\Psi_1\rangle_{A_1B_1C_1} \otimes |\Psi_1\rangle_{A_2B_2C_2} = \left(\frac{1}{\sqrt{2}}\right)^3 [|\Phi^+\rangle_{A_1A_2}|\Phi^+\rangle_{B_1B_2}|\Phi^+\rangle_{C_1C_2} + |\Phi^+\rangle_{A_1A_2}|\Phi^-\rangle_{B_1B_2}|\Phi^-\rangle_{C_1C_2} + |\Phi^-\rangle_{A_1A_2}|\Phi^+\rangle_{B_1B_2}|\Phi^-\rangle_{C_1C_2}$$
$$+ |\Phi^-\rangle_{A_1A_2}|\Phi^-\rangle_{B_1B_2}|\Phi^+\rangle_{C_1C_2} + |\Psi^+\rangle_{A_1A_2}|\Psi^+\rangle_{B_1B_2}|\Psi^+\rangle_{C_1C_2} + |\Psi^+\rangle_{A_1A_2}|\Psi^-\rangle_{B_1B_2}|\Psi^-\rangle_{C_1C_2}$$
$$+ |\Psi^-\rangle_{A_1A_2}|\Psi^+\rangle_{B_1B_2}|\Psi^-\rangle_{C_1C_2} + |\Psi^-\rangle_{A_1A_2}|\Psi^-\rangle_{B_1B_2}|\Psi^+\rangle_{C_1C_2}] \quad (6)$$

$$|\Psi_1\rangle_{A_1B_1C_1} \otimes |\Psi_2\rangle_{A_2B_2C_2} = \left(\frac{1}{\sqrt{2}}\right)^3 [|\Phi^+\rangle_{A_1A_2}|\Phi^+\rangle_{B_1B_2}|\Phi^-\rangle_{C_1C_2} + |\Phi^+\rangle_{A_1A_2}|\Phi^-\rangle_{B_1B_2}|\Phi^+\rangle_{C_1C_2} + |\Phi^-\rangle_{A_1A_2}|\Phi^+\rangle_{B_1B_2}|\Phi^+\rangle_{C_1C_2}$$
$$+ |\Phi^-\rangle_{A_1A_2}|\Phi^-\rangle_{B_1B_2}|\Phi^-\rangle_{C_1C_2} - |\Psi^+\rangle_{A_1A_2}|\Psi^+\rangle_{B_1B_2}|\Psi^-\rangle_{C_1C_2} - |\Psi^+\rangle_{A_1A_2}|\Psi^-\rangle_{B_1B_2}|\Psi^+\rangle_{C_1C_2}$$
$$- |\Psi^-\rangle_{A_1A_2}|\Psi^+\rangle_{B_1B_2}|\Psi^+\rangle_{C_1C_2} - |\Psi^-\rangle_{A_1A_2}|\Psi^-\rangle_{B_1B_2}|\Psi^-\rangle_{C_1C_2}] \quad (7)$$

$$|\Psi_1\rangle_{A_1B_1C_1} \otimes |\Psi_3\rangle_{A_2B_2C_2} = \left(\frac{1}{\sqrt{2}}\right)^3 [|\Psi^+\rangle_{A_1A_2}|\Phi^+\rangle_{B_1B_2}|\Phi^+\rangle_{C_1C_2} + |\Psi^+\rangle_{A_1A_2}|\Phi^-\rangle_{B_1B_2}|\Phi^-\rangle_{C_1C_2} + |\Psi^-\rangle_{A_1A_2}|\Phi^+\rangle_{B_1B_2}|\Phi^-\rangle_{C_1C_2}$$
$$+ |\Psi^-\rangle_{A_1A_2}|\Phi^-\rangle_{B_1B_2}|\Phi^+\rangle_{C_1C_2} + |\Phi^+\rangle_{A_1A_2}|\Psi^+\rangle_{B_1B_2}|\Psi^+\rangle_{C_1C_2} + |\Phi^+\rangle_{A_1A_2}|\Psi^-\rangle_{B_1B_2}|\Psi^-\rangle_{C_1C_2}$$
$$+ |\Phi^-\rangle_{A_1A_2}|\Psi^+\rangle_{B_1B_2}|\Psi^-\rangle_{C_1C_2} + |\Phi^-\rangle_{A_1A_2}|\Psi^-\rangle_{B_1B_2}|\Psi^+\rangle_{C_1C_2}] \quad (8)$$

$$|\Psi_1\rangle_{A_1B_1C_1} \otimes |\Psi_4\rangle_{A_2B_2C_2} = \left(\frac{1}{\sqrt{2}}\right)^3 [|\Psi^+\rangle_{A_1A_2}|\Phi^+\rangle_{B_1B_2}|\Phi^-\rangle_{C_1C_2} + |\Psi^+\rangle_{A_1A_2}|\Phi^-\rangle_{B_1B_2}|\Phi^+\rangle_{C_1C_2} + |\Psi^-\rangle_{A_1A_2}|\Phi^+\rangle_{B_1B_2}|\Phi^+\rangle_{C_1C_2}$$
$$+ |\Psi^-\rangle_{A_1A_2}|\Phi^-\rangle_{B_1B_2}|\Phi^-\rangle_{C_1C_2} - |\Phi^+\rangle_{A_2A_2}|\Psi^+\rangle_{B_1B_2}|\Psi^-\rangle_{C_1C_2} - |\Phi^+\rangle_{A_1A_2}|\Psi^-\rangle_{B_1B_2}|\Psi^+\rangle_{C_1C_2}$$
$$- |\Phi^-\rangle_{A_1A_2}|\Psi^+\rangle_{B_1B_2}|\Psi^+\rangle_{C_1C_2} - |\Phi^-\rangle_{A_1A_2}|\Psi^-\rangle_{B_1B_2}|\Psi^-\rangle_{C_1C_2}] \quad (9)$$

$$|\Psi_1\rangle_{A_1B_1C_1} \otimes |\Psi_5\rangle_{A_2B_2C_2} = \left(\frac{1}{\sqrt{2}}\right)^3 [|\Phi^+\rangle_{A_1A_2}|\Psi^+\rangle_{B_1B_2}|\Phi^+\rangle_{C_1C_2} + |\Phi^+\rangle_{A_1A_2}|\Psi^-\rangle_{B_1B_2}|\Phi^-\rangle_{C_1C_2} + |\Phi^-\rangle_{A_1A_2}|\Psi^+\rangle_{B_1B_2}|\Phi^-\rangle_{C_1C_2}$$
$$+ |\Phi^-\rangle_{A_1A_2}|\Psi^-\rangle_{B_1B_2}|\Phi^+\rangle_{C_1C_2} + |\Psi^+\rangle_{A_1A_2}|\Phi^+\rangle_{B_1B_2}|\Psi^+\rangle_{C_1C_2} + |\Psi^+\rangle_{A_1A_2}|\Phi^-\rangle_{B_1B_2}|\Psi^-\rangle_{C_1C_2}$$
$$+ |\Psi^-\rangle_{A_1A_2}|\Phi^+\rangle_{B_1B_2}|\Psi^-\rangle_{C_1C_2} + |\Psi^-\rangle_{A_1A_2}|\Phi^-\rangle_{B_1B_2}|\Psi^+\rangle_{C_1C_2}] \quad (10)$$

$$|\Psi_1\rangle_{A_1B_1C_1} \otimes |\Psi_6\rangle_{A_2B_2C_2} = \left(\frac{1}{\sqrt{2}}\right)^3 [|\Phi^+\rangle_{A_1A_2}|\Psi^+\rangle_{B_1B_2}|\Phi^-\rangle_{C_1C_2} + |\Phi^+\rangle_{A_1A_2}|\Psi^-\rangle_{B_1B_2}|\Phi^+\rangle_{C_1C_2} + |\Phi^-\rangle_{A_1A_2}|\Psi^+\rangle_{B_1B_2}|\Phi^+\rangle_{C_1C_2}$$
$$+ |\Phi^-\rangle_{A_1A_2}|\Psi^-\rangle_{B_1B_2}|\Phi^-\rangle_{C_1C_2} - |\Psi^+\rangle_{A_1A_2}|\Phi^+\rangle_{B_1B_2}|\Psi^-\rangle_{C_1C_2} - |\Psi^+\rangle_{A_1A_2}|\Phi^-\rangle_{B_1B_2}|\Psi^+\rangle_{C_1C_2}$$
$$- |\Psi^-\rangle_{A_1A_2}|\Phi^+\rangle_{B_1B_2}|\Psi^+\rangle_{C_1C_2} - |\Psi^-\rangle_{A_1A_2}|\Phi^-\rangle_{B_1B_2}|\Psi^-\rangle_{C_1C_2}] \quad (11)$$

$$|\Psi_1\rangle_{A_1B_1C_1} \otimes |\Psi_7\rangle_{A_2B_2C_2} = \left(\frac{1}{\sqrt{2}}\right)^3 [|\Psi^+\rangle_{A_1A_2}|\Psi^+\rangle_{B_1B_2}|\Phi^+\rangle_{C_1C_2} + |\Psi^+\rangle_{A_1A_2}|\Psi^-\rangle_{B_1B_2}|\Phi^-\rangle_{C_1C_2} + |\Psi^-\rangle_{A_1A_2}|\Psi^+\rangle_{B_1B_2}|\Phi^-\rangle_{C_1C_2}$$
$$+ |\Psi^-\rangle_{A_1A_2}|\Psi^-\rangle_{B_1B_2}|\Phi^+\rangle_{C_1C_2} + |\Phi^+\rangle_{A_1A_2}|\Phi^+\rangle_{B_1B_2}|\Psi^+\rangle_{C_1C_2} + |\Phi^+\rangle_{A_1A_2}|\Phi^-\rangle_{B_1B_2}|\Psi^-\rangle_{C_1C_2}$$
$$+ |\Phi^-\rangle_{A_1A_2}|\Phi^+\rangle_{B_1B_2}|\Psi^-\rangle_{C_1C_2} + |\Phi^-\rangle_{A_1A_2}|\Phi^-\rangle_{B_1B_2}|\Psi^+\rangle_{C_1C_2}] \quad (12)$$

$$|\Psi_1\rangle_{A_1B_1C_1} \otimes |\Psi_8\rangle_{A_2B_2C_2} = \left(\frac{1}{\sqrt{2}}\right)^3 [|\Psi^+\rangle_{A_1A_2}|\Psi^+\rangle_{B_1B_2}|\Phi^-\rangle_{C_1C_2} + |\Psi^+\rangle_{A_1A_2}|\Psi^-\rangle_{B_1B_2}|\Phi^+\rangle_{C_1C_2} + |\Psi^-\rangle_{A_1A_2}|\Psi^+\rangle_{B_1B_2}|\Phi^+\rangle_{C_1C_2}$$
$$+ |\Psi^-\rangle_{A_1A_2}|\Psi^-\rangle_{B_1B_2}|\Phi^-\rangle_{C_1C_2} - |\Phi^+\rangle_{A_1A_2}|\Phi^+\rangle_{B_1B_2}|\Psi^-\rangle_{C_1C_2} - |\Phi^+\rangle_{A_1A_2}|\Phi^-\rangle_{B_1B_2}|\Psi^+\rangle_{C_1C_2}$$
$$- |\Phi^-\rangle_{A_1A_2}|\Phi^+\rangle_{B_1B_2}|\Psi^+\rangle_{C_1C_2} - |\Phi^-\rangle_{A_1A_2}|\Phi^-\rangle_{B_1B_2}|\Psi^-\rangle_{C_1C_2}] \quad (13)$$

According to formulas (6)-(13), each result of $A_1A_2$, $B_1B_2$ and $C_1C_2$ after entanglement swapping do only correspond to one initial state among the above eight known initial states. Corresponding to formulas (6)-(13), eight collections composed by different results of $A_1A_2$, $B_1B_2$ and $C_1C_2$ after entanglement swapping are coded as:

$$\{|\Phi^+\rangle_{A_1A_2}|\Phi^+\rangle_{B_1B_2}|\Phi^+\rangle_{C_1C_2}, |\Phi^+\rangle_{A_1A_2}|\Phi^-\rangle_{B_1B_2}|\Phi^-\rangle_{C_1C_2}, |\Phi^-\rangle_{A_1A_2}|\Phi^+\rangle_{B_1B_2}|\Phi^-\rangle_{C_1C_2}, |\Phi^-\rangle_{A_1A_2}|\Phi^-\rangle_{B_1B_2}|\Phi^+\rangle_{C_1C_2},$$
$$|\Psi^+\rangle_{A_1A_2}|\Psi^+\rangle_{B_1B_2}|\Psi^+\rangle_{C_1C_2}, |\Psi^+\rangle_{A_1A_2}|\Psi^-\rangle_{B_1B_2}|\Psi^-\rangle_{C_1C_2}, |\Psi^-\rangle_{A_1A_2}|\Psi^+\rangle_{B_1B_2}|\Psi^-\rangle_{C_1C_2}, |\Psi^-\rangle_{A_1A_2}|\Psi^-\rangle_{B_1B_2}|\Psi^+\rangle_{C_1C_2}\} \to 000 \quad (14)$$

$$\{|\Phi^+\rangle_{A_1A_2}|\Phi^+\rangle_{B_1B_2}|\Phi^-\rangle_{C_1C_2}, |\Phi^+\rangle_{A_1A_2}|\Phi^-\rangle_{B_1B_2}|\Phi^+\rangle_{C_1C_2}, |\Phi^-\rangle_{A_1A_2}|\Phi^+\rangle_{B_1B_2}|\Phi^+\rangle_{C_1C_2}, |\Phi^-\rangle_{A_1A_2}|\Phi^-\rangle_{B_1B_2}|\Phi^-\rangle_{C_1C_2},$$
$$|\Psi^+\rangle_{A_1A_2}|\Psi^+\rangle_{B_1B_2}|\Psi^-\rangle_{C_1C_2}, |\Psi^+\rangle_{A_1A_2}|\Psi^-\rangle_{B_1B_2}|\Psi^+\rangle_{C_1C_2}, |\Psi^-\rangle_{A_1A_2}|\Psi^+\rangle_{B_1B_2}|\Psi^+\rangle_{C_1C_2}, |\Psi^-\rangle_{A_1A_2}|\Psi^-\rangle_{B_1B_2}|\Psi^-\rangle_{C_1C_2}\} \to 001 \quad (15)$$

$$\{|\Psi^+\rangle_{A_1A_2}|\Phi^+\rangle_{B_1B_2}|\Phi^+\rangle_{C_1C_2}, |\Psi^+\rangle_{A_1A_2}|\Phi^-\rangle_{B_1B_2}|\Phi^-\rangle_{C_1C_2}, |\Psi^-\rangle_{A_1A_2}|\Phi^+\rangle_{B_1B_2}|\Phi^-\rangle_{C_1C_2}, |\Psi^-\rangle_{A_1A_2}|\Phi^-\rangle_{B_1B_2}|\Phi^+\rangle_{C_1C_2},$$
$$|\Phi^+\rangle_{A_1A_2}|\Psi^+\rangle_{B_1B_2}|\Psi^+\rangle_{C_1C_2}, |\Phi^+\rangle_{A_1A_2}|\Psi^-\rangle_{B_1B_2}|\Psi^-\rangle_{C_1C_2}, |\Phi^-\rangle_{A_1A_2}|\Psi^+\rangle_{B_1B_2}|\Psi^-\rangle_{C_1C_2}, |\Phi^-\rangle_{A_1A_2}|\Psi^-\rangle_{B_1B_2}|\Psi^+\rangle_{C_1C_2}\} \to 010 \quad (16)$$

$$\{|\Psi^+\rangle_{A_1A_2}|\Phi^+\rangle_{B_1B_2}|\Phi^-\rangle_{C_1C_2}, |\Psi^+\rangle_{A_1A_2}|\Phi^-\rangle_{B_1B_2}|\Phi^+\rangle_{C_1C_2}, |\Psi^-\rangle_{A_1A_2}|\Phi^+\rangle_{B_1B_2}|\Phi^+\rangle_{C_1C_2}, |\Psi^-\rangle_{A_1A_2}|\Phi^-\rangle_{B_1B_2}|\Phi^-\rangle_{C_1C_2},$$
$$|\Phi^+\rangle_{A_2A_2}|\Psi^+\rangle_{B_1B_2}|\Psi^-\rangle_{C_1C_2}, |\Phi^+\rangle_{A_1A_2}|\Psi^-\rangle_{B_1B_2}|\Psi^+\rangle_{C_1C_2}, |\Phi^-\rangle_{A_1A_2}|\Psi^+\rangle_{B_1B_2}|\Psi^+\rangle_{C_1C_2}, |\Phi^-\rangle_{A_1A_2}|\Psi^-\rangle_{B_1B_2}|\Psi^-\rangle_{C_1C_2}\} \to 011 \quad (17)$$

$$\{|\Phi^+\rangle_{A_1A_2}|\Psi^+\rangle_{B_1B_2}|\Phi^+\rangle_{C_1C_2}, |\Phi^+\rangle_{A_1A_2}|\Psi^-\rangle_{B_1B_2}|\Phi^-\rangle_{C_1C_2}, |\Phi^-\rangle_{A_1A_2}|\Psi^+\rangle_{B_1B_2}|\Phi^-\rangle_{C_1C_2}, |\Phi^-\rangle_{A_1A_2}|\Psi^-\rangle_{B_1B_2}|\Phi^+\rangle_{C_1C_2},$$
$$|\Psi^+\rangle_{A_1A_2}|\Phi^+\rangle_{B_1B_2}|\Psi^+\rangle_{C_1C_2}, |\Psi^+\rangle_{A_1A_2}|\Phi^-\rangle_{B_1B_2}|\Psi^-\rangle_{C_1C_2}, |\Psi^-\rangle_{A_1A_2}|\Phi^+\rangle_{B_1B_2}|\Psi^-\rangle_{C_1C_2}, |\Psi^-\rangle_{A_1A_2}|\Phi^-\rangle_{B_1B_2}|\Psi^+\rangle_{C_1C_2}\} \to 100 \quad (18)$$

$$\{|\Phi^+\rangle_{A_1A_2}|\Psi^+\rangle_{B_1B_2}|\Phi^-\rangle_{C_1C_2}, |\Phi^+\rangle_{A_1A_2}|\Psi^-\rangle_{B_1B_2}|\Phi^+\rangle_{C_1C_2}, |\Phi^-\rangle_{A_1A_2}|\Psi^+\rangle_{B_1B_2}|\Phi^+\rangle_{C_1C_2}, |\Phi^-\rangle_{A_1A_2}|\Psi^-\rangle_{B_1B_2}|\Phi^-\rangle_{C_1C_2},$$
$$|\Psi^+\rangle_{A_1A_2}|\Phi^+\rangle_{B_1B_2}|\Psi^-\rangle_{C_1C_2}, |\Psi^+\rangle_{A_1A_2}|\Phi^-\rangle_{B_1B_2}|\Psi^+\rangle_{C_1C_2}, |\Psi^-\rangle_{A_1A_2}|\Phi^+\rangle_{B_1B_2}|\Psi^+\rangle_{C_1C_2}, |\Psi^-\rangle_{A_1A_2}|\Phi^-\rangle_{B_1B_2}|\Psi^-\rangle_{C_1C_2}\} \to 101 \quad (19)$$

$$\{|\Psi^+\rangle_{A_1A_2}|\Psi^+\rangle_{B_1B_2}|\Phi^+\rangle_{C_1C_2}, |\Psi^+\rangle_{A_1A_2}|\Psi^-\rangle_{B_1B_2}|\Phi^-\rangle_{C_1C_2}, |\Psi^-\rangle_{A_1A_2}|\Psi^+\rangle_{B_1B_2}|\Phi^-\rangle_{C_1C_2}, |\Psi^-\rangle_{A_1A_2}|\Psi^-\rangle_{B_1B_2}|\Phi^+\rangle_{C_1C_2},$$
$$|\Phi^+\rangle_{A_1A_2}|\Phi^+\rangle_{B_1B_2}|\Psi^+\rangle_{C_1C_2}, |\Phi^+\rangle_{A_1A_2}|\Phi^-\rangle_{B_1B_2}|\Psi^-\rangle_{C_1C_2}, |\Phi^-\rangle_{A_1A_2}|\Phi^+\rangle_{B_1B_2}|\Psi^-\rangle_{C_1C_2}, |\Phi^-\rangle_{A_1A_2}|\Phi^-\rangle_{B_1B_2}|\Psi^+\rangle_{C_1C_2}\} \to 110 \quad (20)$$

$$\{|\Psi^+\rangle_{A_1A_2}|\Psi^+\rangle_{B_1B_2}|\Phi^-\rangle_{C_1C_2}, |\Psi^+\rangle_{A_1A_2}|\Psi^-\rangle_{B_1B_2}|\Phi^+\rangle_{C_1C_2}, |\Psi^-\rangle_{A_1A_2}|\Psi^+\rangle_{B_1B_2}|\Phi^+\rangle_{C_1C_2}, |\Psi^-\rangle_{A_1A_2}|\Psi^-\rangle_{B_1B_2}|\Phi^-\rangle_{C_1C_2},$$

$$\left|\Phi^+\right\rangle_{A_1A_2}\left|\Phi^+\right\rangle_{B_1B_2}\left|\Psi^-\right\rangle_{C_1C_2}, \left|\Phi^+\right\rangle_{A_1A_2}\left|\Phi^-\right\rangle_{B_1B_2}\left|\Psi^+\right\rangle_{C_1C_2}, \left|\Phi^-\right\rangle_{A_1A_2}\left|\Phi^+\right\rangle_{B_1B_2}\left|\Psi^+\right\rangle_{C_1C_2}, \left|\Phi^-\right\rangle_{A_1A_2}\left|\Phi^-\right\rangle_{B_1B_2}\left|\Psi^-\right\rangle_{C_1C_2}\} \to 111 \quad (21)$$

Further extending the initial state from $\left|\Psi_1\right\rangle$ to other seven GHZ-states $\left|\Psi_k\right\rangle$ ($k=2,\cdots,8$), all the result collections composed by different results of $A_1A_2$, $B_1B_2$ and $C_1C_2$ after entanglement swapping are listed in Table1. Take the initial state of particles $A_1$, $B_1$, $C_1$ to be $\left|\Psi_2\right\rangle_{A_1B_1C_1}^{001}$ and the initial state of particles $A_2$, $B_2$, $C_2$ to be $\left|\Psi_6\right\rangle_{A_2B_2C_2}^{101}$ for example. The superscript in $\left|\Psi_2\right\rangle_{A_1B_1C_1}^{001}$ denote that $\left|\Psi_2\right\rangle_{A_1B_1C_1}^{001}$ can be obtained by performing $U_2$ on the first and second particles of $\left|\Psi_1\right\rangle_{A_1B_1C_1}^{000}$, while 100 denotes that the result collection composed by $A_1A_2$, $B_1B_2$ and $C_1C_2$ after entanglement swapping from $\left|\Psi_2\right\rangle_{A_1B_1C_1}^{001}$ and $\left|\Psi_6\right\rangle_{A_2B_2C_2}^{101}$ corresponds to formula (18).

Table 1. The result collections of entanglement swapping between any two GHZ states
(The superscript denotes the codes of $U_k$ and the subscript denotes the particles in the GHZ states)

| | $\left|\Psi_1\right\rangle_{A_2B_2C_2}^{000}$ | $\left|\Psi_2\right\rangle_{A_2B_2C_2}^{001}$ | $\left|\Psi_3\right\rangle_{A_2B_2C_2}^{010}$ | $\left|\Psi_4\right\rangle_{A_2B_2C_2}^{011}$ | $\left|\Psi_5\right\rangle_{A_2B_2C_2}^{100}$ | $\left|\Psi_6\right\rangle_{A_2B_2C_2}^{101}$ | $\left|\Psi_7\right\rangle_{A_2B_2C_2}^{110}$ | $\left|\Psi_8\right\rangle_{A_2B_2C_2}^{111}$ |
|---|---|---|---|---|---|---|---|---|
| $\left|\Psi_1\right\rangle_{A_1B_1C_1}^{000}$ | 000 | 001 | 010 | 011 | 100 | 101 | 110 | 111 |
| $\left|\Psi_2\right\rangle_{A_1B_1C_1}^{001}$ | 001 | 000 | 011 | 010 | 101 | 100 | 111 | 110 |
| $\left|\Psi_3\right\rangle_{A_1B_1C_1}^{010}$ | 010 | 011 | 000 | 001 | 110 | 111 | 100 | 101 |
| $\left|\Psi_4\right\rangle_{A_1B_1C_1}^{011}$ | 011 | 010 | 001 | 000 | 111 | 110 | 101 | 100 |
| $\left|\Psi_5\right\rangle_{A_1B_1C_1}^{100}$ | 100 | 101 | 110 | 111 | 000 | 001 | 010 | 011 |
| $\left|\Psi_6\right\rangle_{A_1B_1C_1}^{101}$ | 101 | 100 | 111 | 110 | 001 | 000 | 011 | 010 |
| $\left|\Psi_7\right\rangle_{A_1B_1C_1}^{110}$ | 110 | 111 | 100 | 101 | 010 | 011 | 000 | 001 |
| $\left|\Psi_8\right\rangle_{A_1B_1C_1}^{111}$ | 111 | 110 | 101 | 100 | 011 | 010 | 001 | 000 |

## 3 Quantum steganography protocol

Our quantum steganography protocol integrates the original QSDC adopting dense coding of GHZ states, which is inspired by Ref.[7] and Ref.[9], and entanglement swapping of GHZ states together. In the schemes of Ref.[7] and Ref.[9], all particles are eventually transmitted from one communication party to the other communication party. However, in this original QSDC, the third particle from each GHZ triplet is always kept intact in one communication party's hand, while the first particle and the second particle are transmitted between two communication parties. The basic idea of the original QSDC is that: (1) Bob keeps the sequence of the third particle from each GHZ triplet in hand, and sends the sequence of the first particle and the sequence of the second particle to Alice one by one. In order to ensure security, eavesdropping detection is implemented in each transmission; (2) According to the bits sequence of information, Alice then adopts dense coding of GHZ states to perform unitary operations on the two sequences. Afterward, Alice sends the two encoded sequences back to Bob; (3) Finally, Bob implements GHZ-basis measurement on each GHZ triplet to recover the information. Due to the dense coding of GHZ states, the original QSDC can transmit three bits information per round communication. Now, we demonstrate our quantum steganography protocol in detail as follows.

S1) Bob prepares a large number ($n$) of $\left|\Psi_1\right\rangle_{ABC}$. Let $G_A$, $G_B$ and $G_C$ denote the particle groups of $A$, $B$ and $C$, respectively. Accordingly, $G_A = [A_1, A_2, \cdots, A_n]$, $G_B = [B_1, B_2, \cdots, B_n]$ and $G_C = [C_1, C_2, \cdots, C_n]$, where the subscript denotes the number of GHZ state.

S2) Bob sends $G_A$ and $G_B$ to Alice by quantum channel in two steps: (a) Bob sends $G_A$ to Alice while keeping $G_B$ and $G_C$ for himself. For eavesdropping detection, Alice selects a large enough subgroup from $G_A$, and chooses randomly a measurement basis $Z$-basis ($\left|0\right\rangle, \left|1\right\rangle$) or $X$-basis ($\left|+\right\rangle, \left|-\right\rangle$) to measure particle $A$ in subgroup from $G_A$. Alice publishes her measurement basis and measurement results to Bob. After obtaining Alice's results, Bob measures particle $B$ from the corresponding subgroup of $G_B$ and particle $C$ from the corresponding subgroup of $G_C$ under the same measurement basis. According to formulas (22), by comparing with Alice's measurement results, Bob can know whether there is an eavesdropping or not. If the channel is safe, their measurement results are highly correlated. When Alice and Bob perform $Z$-basis, Bob's measurement result should be $\left|0\right\rangle\left|0\right\rangle$ ($\left|1\right\rangle\left|1\right\rangle$) if Alice gets the measurement result $|0\rangle$ ( $|1\rangle$ ). In addition, when Alice and Bob perform $X$-basis, Bob's measurement result should be $|+\rangle|+\rangle$ or $|-\rangle|-\rangle$ ( $|+\rangle|-\rangle$ or $|-\rangle|+\rangle$ )if Alice gets the measurement result $|+\rangle$ ( $|-\rangle$ ). Then, if Bob confirms that there is an eavesdropping, they abort the communication; otherwise, they enter the following step (b); (b) Bob sends $G_B$ to Alice while keeping $G_C$ for himself. For eavesdropping detection, Alice selects a large enough subgroup from $G_A$ and a large enough corresponding subgroup from $G_B$, and chooses randomly a measurement basis $Z$-basis $(|0\rangle,|1\rangle)$ or $X$-basis $(|+\rangle,|-\rangle)$ to measure particle $A$ and particle $B$. Alice publishes her measurement basis and measurement results to Bob. After obtaining Alice's results, Bob measures particle $C$ from the corresponding subgroup of $G_C$ under the same measurement basis. According to formulas (22), by comparing with Alice's measurement results, Bob can know whether there is an eavesdropping or not. If the channel is safe, their measurement results are highly correlated. Then, if Bob confirms that there is an eavesdropping, they abort the communication; otherwise, they enter information transmission mode S3.

$$|\Psi_1\rangle_{ABC} = \frac{1}{\sqrt{2}}(|000\rangle_{ABC}+|111\rangle_{ABC}) = \frac{1}{2}[|+\rangle_A(|+\rangle_B|+\rangle_C+|-\rangle_B|-\rangle_C)+|-\rangle_A(|+\rangle_B|-\rangle_C+|-\rangle_B|+\rangle_C)]$$
$$= \frac{1}{2}[(|+\rangle_A|+\rangle_B+|-\rangle_A|-\rangle_B)|+\rangle_C+(|+\rangle_A|-\rangle_B+|-\rangle_A|+\rangle_B)|-\rangle_C] \tag{22}$$

S3) Information transmission mode: (a) According to the bits sequence of information, Alice performs $U_k$ on the pairs of particles in $G_A$ and $G_B$ (Note that after performed with $U_k$, $G_A$ and $G_B$ turns to be $G'_A$ and $G'_B$, respectively. Although no unitary operations have been performed on the particle in $G_C$, for consistence, $G'_C$ is still used to represent the original $G_C$. Accordingly, $G'_C$ is the same as $G_C$.); (b) According to secret messages, Alice chooses four particles $A'_m$, $A'_{m+1}$, $B'_m$, $B'_{m+1}$ from $G'_A$ and $G'_B$, respectively, and enters secret messages hiding mode; (c) Alice sends $G'_A$ and $G'_B$ back to Bob by quantum channel.

S4) Secret messages hiding mode: (a) According to secret messages, Alice chooses four particles $A'_m$, $A'_{m+1}$, $B'_m$, $B'_{m+1}$ from $G'_A$ and $G'_B$, respectively, where the subscript $m$ represents the position of the particle $A'_m$ in $G'_A$ and the position of the particle $B'_m$ in $G'_B$. The value of $m$ must satisfy the consistent condition, which means two GHZ states formed by $A'_{m-1}B'_{m-1}C'_{m-1}$ and $A'_mB'_mC'_m$ must be consistent with the secret messages, as shown in Table 1(An appropriate $m$ can be decided by Alice in advance before sending $m$ to Bob by implementing QSDC, QKD or one-time pad through classical channel[23].); (b) By performing the same $U_k$ on $A'_{m+1}$ and $B'_{m+1}$ in advance, $A'_{m+1}B'_{m+1}C'_{m+1}$ can copy the information carried by $A'_{m-1}B'_{m-1}C'_{m-1}$. Consequently, $A'_{m+1}B'_{m+1}C'_{m+1}$ doesn't normally transmit information but acts as an auxiliary GHZ state to help hide secret messages.

S5) Secret messages decoding mode: (a) Bob gets the value of $m$; (b) Bob performs GHZ-basis measurement on $A'_{m-1}B'_{m-1}C'_{m-1}$ to recover the information; (c) Bob performs Bell-basis measurement on $A'_mA'_{m+1}$, $B'_mB'_{m+1}$ and $C'_mC'_{m+1}$, respectively; (d) Then, Bob decodes secret messages sent by Alice, according to formulas (14)-(21). Moreover, through the decoded secret messages and the state of $A'_{m-1}B'_{m-1}C'_{m-1}$, Bob can also recover the information carried by $A'_mB'_mC'_m$, according to Table 1.

We use an example to further explain secret messages hiding mode S4 and secret messages decoding S5 in our quantum steganography protocol. Assuming that the secret messages Alice wants to send Bob are **100**, and the information sequence $\cdots 000_{100}\cdots 001_{101}\cdots 010_{110}\cdots 011_{111}\cdots 100_{000}\cdots 101_{001}\cdots 110_{010}\cdots 111_{011}\cdots$ is generated by Alice (The information is divided by six bits, since two $U_k$ denote six bits information). Assume that the group numbers of $000_{100}$, $001_{101}$, $010_{110}$, $011_{111}$, $100_{000}$, $101_{001}$, $110_{010}$, $111_{011}$ in the information sequence are No.7, 10, 13, 16, 20, 25, 28 and 32, respectively. In S3, Alice can make $m = 7, 10, 13, 16, 20, 25, 28$ or 32 to satisfy the consistency in Table 1. If $m = 7$, $A'_6B'_6C'_6$ will be $|\Psi_1\rangle$ and $A'_7B'_7C'_7$ will be $|\Psi_5\rangle$. Accordingly, the secret messages **100** are transmitted by entanglement swapping between $A'_7B'_7C'_7$ and $A'_8B'_8C'_8$. The secret messages **100** can also be transmitted in the same way, if $m = 10, 13, 16, 20, 25, 28$ or 32. Note that, $A'_8B'_8C'_8$ can't be used to transmit the information like other normal GHZ states, and acts as an auxiliary GHZ state to help hide secret messages. In S6, Bob obtains the value of $m$ at first. Afterward, Bob performs GHZ-basis measurement on $A'_6B'_6C'_6$. Then, Bob performs Bell-basis measurement on $A'_7A'_8$, $B'_7B'_8$ and $C'_7C'_8$, respectively. According to formulas (14)-(21), Bob can decode that the secret messages are **100**. Then, according to the state of $A'_6B'_6C'_6$ ( $|\Psi_1\rangle$ ) and the secret messages **100**, according to Table 1, Bob can easily know that the information carried by $A'_7B'_7C'_7$ is **100**.

## 4 Analysis

## 4.1 Capacity

In the above quantum steganography protocol, three bits secret messages are transmitted by entanglement swapping between $A'_{m+1}B'_{m+1}C'_{m+1}$ and $A'_{m}B'_{m}C'_{m}$. In addition, $A'_{m+1}B'_{m+1}C'_{m+1}$ copies the information carried by $A'_{m-1}B'_{m-1}C'_{m-1}$ and acts as an auxiliary GHZ state to help hide secret messages. Accordingly, the information carried by $A'_{m}B'_{m}C'_{m}$ is recovered, while $A'_{m+1}B'_{m+1}C'_{m+1}$ is consumed. Moreover, it is clear that three bits secret messages can be transmitted by eight different kinds of initial states. For example, according to Table 1, **100** can be transmitted by eight different kinds of initial states $\{|\Psi_1\rangle^{000}_{A_1B_1C_1}, |\Psi_5\rangle^{100}_{A_2B_2C_2}\}$, $\{|\Psi_2\rangle^{001}_{A_1B_1C_1}, |\Psi_6\rangle^{101}_{A_2B_2C_2}\}$, $\{|\Psi_3\rangle^{010}_{A_1B_1C_1}, |\Psi_7\rangle^{110}_{A_2B_2C_2}\}$, $\{|\Psi_4\rangle^{011}_{A_1B_1C_1}, |\Psi_8\rangle^{111}_{A_2B_2C_2}\}$, $\{|\Psi_5\rangle^{100}_{A_1B_1C_1}, |\Psi_1\rangle^{000}_{A_2B_2C_2}\}$, $\{|\Psi_6\rangle^{101}_{A_1B_1C_1}, |\Psi_2\rangle^{001}_{A_2B_2C_2}\}$, $\{|\Psi_7\rangle^{110}_{A_1B_1C_1}, |\Psi_3\rangle^{010}_{A_2B_2C_2}\}$, $\{|\Psi_8\rangle^{111}_{A_1B_1C_1}, |\Psi_4\rangle^{011}_{A_2B_2C_2}\}$. After coding these eight different kinds of initial states by formulas (23), the capacity of quantum channel in the above quantum steganography protocol can be increased to six bits. Consequently, its capacity of quantum channel is six times that of Ref.[19], Ref.[20], Ref.[21]or Ref.[22], and one and half times that of Ref.[23]. The reason why its capacity of quantum channel is larger than that of Ref.[23] lies in two points: (a) Entanglement swapping between two GHZ states transmits three bits in our quantum steganography protocol, while entanglement swapping between two Bell states only transmits two bits in Ref.[23]; (b) Each three bits secret messages corresponds to eight different kinds of initial states in our quantum steganography protocol, while each two bits secret messages corresponds to four different kinds of initial states in Ref.[23].

$$000100 \to 000, 001101 \to 001, 010110 \to 010, 011111 \to 011, 100000 \to 100, 101001 \to 101, 110010 \to 110, 111011 \to 111 \quad (23)$$

Based on the above analysis, our quantum steganography protocol can transmit six bits per round covert communication. In fact, our quantum steganography protocol transmits secret messages by building up the hidden channel within the original QSDC. However, the original QSDC only transmits three bits per round covert communication. Thus, the transmission efficiency of our quantum steganography protocol is twice that of the original QSDC. It can be concluded that the super quantum channel integrating the original quantum channel of QSDC and the hidden channel in our quantum steganography protocol can enlarge the capacity of quantum channel. It is also possible to apply the idea of our quantum steganography protocol into QSS and QKD based on GHZ states to enlarge the transmission efficiency of the original quantum channel.

## 4.2 Imperceptibility

In our quantum steganography protocol, the choice of $m$ is not arbitrary for Alice, since the value of $m$ must satisfy the consistent condition with respect to $A'_{m-1}B'_{m-1}C'_{m-1}$, $A'_{m}B'_{m}C'_{m}$ and secret messages. Consequently, the imperceptibility mainly depends on the difficulty of knowing $m$ by Eve. As pointed out in Ref.[23], choosing $m$ can still be treated as arbitrary behave for Eve, since both information and secret messages can be regarded to be random or pseudo-random. If information or secret messages doesn't distribute randomly in advance, pseudo-random sequence encryption can be adopted to make its distribution randomized.

For example, if the secret messages Alice wants to send Bob are **100**, in order to choose $m$, Alice needs to find out all the group numbers of "**000**$_{100}$", "**001**$_{101}$", "**010**$_{110}$", "**011**$_{111}$", "**100**$_{000}$", "**101**$_{001}$", "**110**$_{010}$", "**111**$_{011}$" in the information sequence. Accordingly, the GHZ states $A'_{m-1}B'_{m-1}C'_{m-1}$ and $A'_{m}B'_{m}C'_{m}$ will be "$|\Psi_1\rangle|\Psi_5\rangle$", "$|\Psi_2\rangle|\Psi_6\rangle$", "$|\Psi_3\rangle|\Psi_7\rangle$", "$|\Psi_4\rangle|\Psi_8\rangle$", "$|\Psi_5\rangle|\Psi_1\rangle$", "$|\Psi_6\rangle|\Psi_2\rangle$", "$|\Psi_7\rangle|\Psi_3\rangle$" and "$|\Psi_8\rangle|\Psi_4\rangle$", respectively. If the information distributes evenly, probability of "**000**$_{100}$", "**001**$_{101}$", "**010**$_{110}$", "**011**$_{111}$", "**100**$_{000}$", "**101**$_{001}$", "**110**$_{010}$", "**111**$_{011}$" will be $1/64$, respectively. Consequently, their total probability is $1/8$. It is straightforward if the secret messages are **000,001,010,011,101,110** or **111**. Therefore, the probability distributions of information and secret messages will make $m$'s uncertainty best, according to Shannon's information theory, as pointed out in Ref.[23]. Consequently, choosing $m$ can be regarded to be random for Eve. It means that the imperceptibility of our quantum steganography protocol is good.

## 4.3 Security

The security of our protocol can be proved via the security of the original QSDC. The original QSDC uses GHZ states, and its security is similar to the scheme using Bell states in Ref.[7]. The security of the original QSDC is based on the security for the transmission of $G_A$ and $G_B$ from Bob to Alice.

Now we analyze the security for the transmission of $G_A$ against the entanglement-and-measurement attack at first. According to Stinespring dilation theorem, the eavesdropping of Eve can be realized by a unitary operation $\hat{E}$ on a larger Hilbert space, $|x, E\rangle \equiv |x\rangle|E\rangle$. Therefore, the state of the composite system will be

$$|\psi\rangle = \frac{1}{\sqrt{2}}\left[\left(\alpha_1|0\rangle|\varepsilon_{00}\rangle + \beta_1|1\rangle|\varepsilon_{01}\rangle\right)|00\rangle + \left(\beta'_1|0\rangle|\varepsilon_{10}\rangle + \alpha'_1|1\rangle|\varepsilon_{11}\rangle\right)|11\rangle\right] \quad (24)$$

where $\varepsilon_{00}, \varepsilon_{01}, \varepsilon_{10}, \varepsilon_{11}$ are Eve's states, and $\hat{E} = \begin{pmatrix} \alpha_1 & \beta'_1 \\ \beta_1 & \alpha'_1 \end{pmatrix}$ is Eve's probe operator. Since $\hat{E}$ is a unitary operator, we can obtain that the error rate introduced by Eve's eavesdropping on $G_A$ is $\tau_1 = |\beta_1|^2 = |\beta'_1|^2 = 1 - |\alpha_1|^2 = 1 - |\alpha'_1|^2$. The security for the transmission of $G_B$ against the entanglement-and-measurement attack can be analyzed in a similar way as above. After Eve eavesdrops $G_B$ before the second checking, the state of the composite system will be

$$|\psi\rangle = \frac{1}{\sqrt{2}}\left[|0\rangle\left(\alpha_2|0\rangle|\varepsilon_{00}\rangle + \beta_2|1\rangle|\varepsilon_{01}\rangle\right)|0\rangle + |1\rangle\left(\beta_2'|0\rangle|\varepsilon_{10}\rangle + \alpha_2'|1\rangle|\varepsilon_{11}\rangle\right)|1\rangle\right] \quad (25)$$

We can obtain that the error rate introduced by Eve's eavesdropping on $G_B$ will be $\tau_2 = |\beta_2|^2 = |\beta_2'|^2 = 1-|\alpha_2|^2 = 1-|\alpha_2'|^2$.

. Without loss of generality, we turn to analyze the measure-resend attack on $G_A$. Eve intercepts particle $A$ in $G_A$, measures it in $Z$-basis or $X$-basis, and resends its measurement result to Alice. In the first case, Eve performs $Z$-basis measurement. The state of the whole system will collapse to $|000\rangle$ or $|111\rangle$ each with probability $1/2$. Take the state to be $|000\rangle_{ABC}$ for example. Accordingly, Eve resends $|0\rangle_A$ to Alice. Then, if Alice performs $Z$-basis measurement to check eavesdropping, no error will be introduced by Eve. If Alice performs $X$-basis measurement, the state will collapse to $|opq\rangle_{ABC}$ $(o,p,q=+,-)$ each with probability $1/8$. According to formulas (22), the error rate introduced by Eve will be $50\%$. Therefore, the total error rate in this case is $25\%$. In the second case, Eve performs $X$-basis. Then, the total error rate in this case is $37.5\%$. Therefore, the random $Z$-basis or $X$-basis measurement guarantees that Eve's attack can be found out by eavesdropping check.

Further consider the influence caused by leakage of $m$. Assume that Eve not only obtains $m$ but also gets $A'_m A'_{m+1}$ and $B'_m B'_{m+1}$ through some advance eavesdropping attacks (It is possible for Eve to eavesdrop $A'_m A'_{m+1}$ and $B'_m B'_{m+1}$, since $A$ particle and $B$ particle are travel particles). However, Eve still can't get secret messages according to formulas (14)-(21), because only knowing $A'_m A'_{m+1}$ and $B'_m B'_{m+1}$ is not enough to decode secret messages.

## 5  Discussions and Conclusions

As analyzed in section 4.1, the capacity of super quantum channel in our quantum steganography protocol achieves six bits, and is twice that of the original QSDC. The reason lies in that the super quantum channel is formed by building up the hidden channel within the original QSDC. However, the hidden channel works at the cost of transmitting $m$. Transmitting $m$ means to transmit $\log_2 m$ bits. If $m$ is great enough, transmitting $m$ may consume more bits than secret messages. Fortunately, as pointed out in Ref.[23], since $m$ can be decided and transmitted by implementing QSDC, QKD or one-time pad through classical channel in advance, it is unnecessary to overemphasize the cost of transmitting $m$. Furthermore, since $m$ and secret messages tend to have different security level, it is reasonable to consume a certain resource to achieve covert communication for secret messages.

To sum up, a quantum steganography protocol with large payload is proposed, based on dense coding and entanglement swapping of GHZ states. Its super quantum channel is formed by building up the hidden channel within the original QSDC. Based on the original QSDC, secrets messages are transmitted by integrating dense coding of GHZ states and entanglement swapping of GHZ states. It can transmit six bits per round covert communication, much higher than the previous quantum steganography protocols. Since the information and secrets messages can be regarded to be random or pseudo-random, its imperceptibility is good. Moreover, its security is proved to be reliable.

**Acknowledgements**

Funding by the National Natural Science Foundation of China (Grant No. 60972071), and the Natural Science Foundation of Zhejiang Province (Grant Nos. Y6100421 and LQ12F02012) is gratefully acknowledged.